\documentclass[a4paper,fleqn,usenatbib]{mnras}

\usepackage{newtxtext,newtxmath}

\usepackage[T1]{fontenc}
\usepackage{ae,aecompl}

\usepackage{graphicx}   
\usepackage{amsmath}    
\usepackage{amssymb}    
\usepackage{makecell}   
\usepackage{threeparttable} 

\title[Correlation between the continuum and BALs]{Correlation between the variation of the ionizing continuum and broad absorption lines}

\author[W.-J. Lu et al.]{
Wei-Jian Lu,$^{1}$\footnotemark[1]
Ying-Ru Lin,$^{1}$
Yi-Ping Qin$^{2,3}$
\\
$^{1}$School of Information Engineering, Baise University, Baise 533000, China\\
$^{2}$School of Materials Science and Engineering, Baise University, Baise 533000, China\\
$^{3}$Center for Astrophysics, Guangzhou University, Guangzhou 510006, China}

\date{Accepted 2017 October 24. Received 2017 September 22; in original form 2017 June 29}

\pubyear{2017}

\begin{document}
\label{secondpage}
\pagerange{\pageref{secondpage}--\pageref{lastpage}}
\maketitle

\begin{abstract}
In this Letter, we present an analysis of the relation between the variability of broad absorption lines (BALs) and that of the continuum. Our sample is multi-epoch observations of 483 quasars by the Sloan Digital Sky Survey-I/II/III (SDSS-I/II/III). We derive the fractional flux variations of the continuum and fractional equivalent width (EW) variations for \ion{C}{iv} and \ion{Si}{iv} BALs, and explore the correlations between the three. Our results reveal moderate anticorrelations with high significance level between the fractional flux variations of the continuum and fractional EW variations for both \ion{C}{iv} and \ion{Si}{iv} BALs. We also prove a significant positive correlation between the fractional EW variations for \ion{C}{iv} and \ion{Si}{iv} BALs, which is in agreement with several previous studies. Our discoveries can serve as evidence for the idea: Change of an ionizing continuum is the primary driver of BAL variability. 

\end{abstract}
%

\footnotetext[1]{E-mail:william\_lo@qq.com}



\section{Introduction}
Previous observations have long found that broad absorption line (BAL; with absorption widths >2000 $\rm km~s^{-1}$; Weymann et al. 1991) variability on time-scales of months to years (in the quasar rest frame) is common (e.g., Foltz et al. 1987; Smith \& Penston 1988; Barlow, Junkkarinen \& Burbidge 1989). BAL variability may be caused by mechanisms of either changes in the coverage fraction, due, for example, to gas transverse motion (e.g. Hamann et al. 2008; Shi et al. 2016); or changes in ionization states of the absorption gas (e.g., Crenshaw, Kraemer \& George 2003, and references therein).
 
Some studies have attempted to identify the leading mechanism of BAL variability, by analysing the relation between variation of BALs and that of the continuum/emission, and relation between variation of different ions. As a result, most of them have not found obvious relation between the BAL and the continuum variability, which seem to mean that the changes in ionization state is not the dominant mechanism driving BAL variability (Gibson et al. 2008; Wildy, Goad \& Allen 2014; Vivek et al. 2014). Gibson et al. (2008) made the above conclusion based on 13 BAL QSOs observed by both the Large Bright Quasar Survey (LBQS) and the Sloan Digital Sky Survey (SDSS) separating by 3--6 rest-frame years. Wildy et al. (2014) also have not found correlation between quasar luminosity and BAL variability, according to a variability study of 59 \ion{C}{iv} and 38 \ion{Si}{iv} BALs in 50 BAL quasars. Similar result has been presented by Vivek et al. (2014) based on multi-epoch spectroscopic data of 22 low-ionization BAL quasars that contain \ion{Mg}{ii} and \ion{Al}{iii} BALs.
 
However, correlations between the variation in different ions of the same broad absorption component do have been found (Filiz et al. 2013; Wildy et al. 2014). For instance, using the multi-epoch SDSS spectra of 291 quasars containing 428 variable \ion{C}{iv} and 235 \ion{Si}{iv} BALs, Filiz et al. (2013) have found a strong correlation (at a significance level of $\textgreater $99.9 per cent) between the variations in equivalent widths (EWs) of \ion{C}{iv} and \ion{Si}{iv} BALs corresponding in velocity. Additionally, Wang et al. (2015) have presented a discovery of the coordinated variability between BALs and the ionizing continuum, though only based on a qualitative analysis. Furthermore, He et al. (2017) estimated statistically that BAL variability in at least 80 per cent quasars is mainly caused by the variation in the ionizing continuum, based on a large sample of multi-observed BAL quasars from SDSS-I/II/III (Shen et al. 2011; P\^aris et al. 2017).
\begin{figure*}
\includegraphics[width=2\columnwidth]{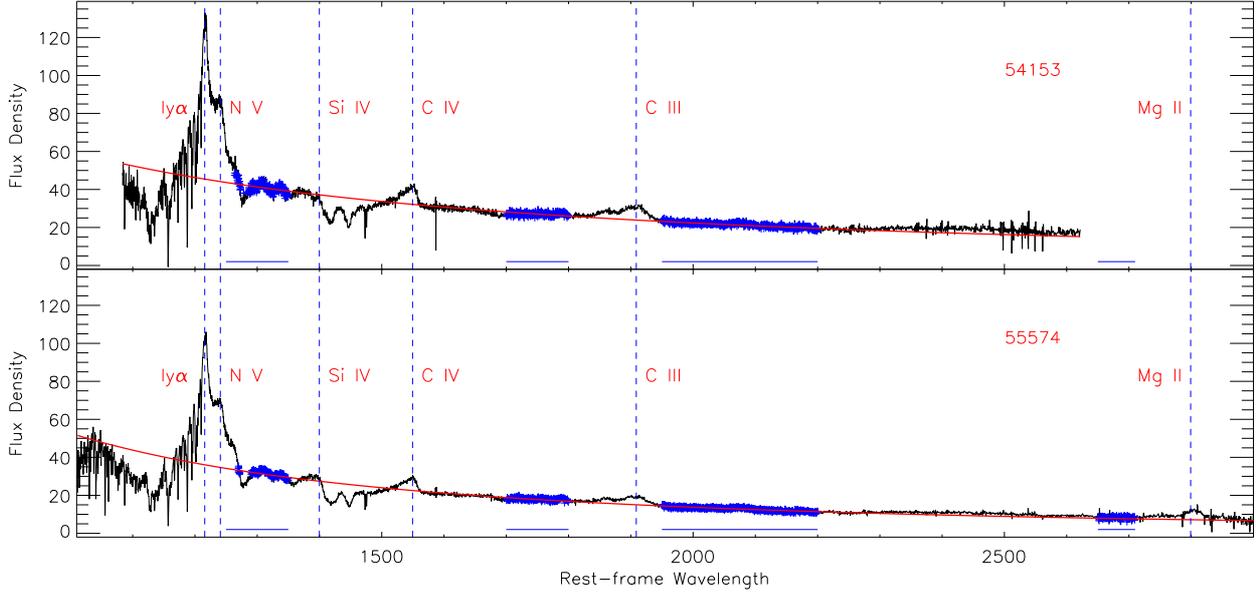}
\caption{An example of the power-law continuum fit (of SDSS J081026.42+101041.0). The top and bottom panels represent the two epochs (the SDSS MJDs are labelled in the top right-hand corner of each panel). The observed flux density is in units of $10^{-17}$ erg cm$^{-2}$ s$^{-1}$\AA$^{-1}$ and the rest-frame wavelength is in units of \AA. The blue horizontal lines are the RLF windows, and the blue crosses are the pixels that used to fit the power-law continuum. The main emissions are marked out with blue vertical dotted lines.}
    \label{fig.1}
\end{figure*}

\begin{figure}
\includegraphics[width=\columnwidth]{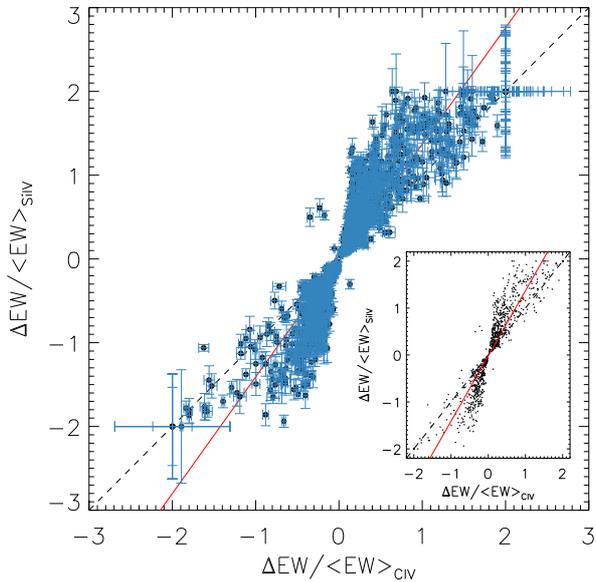}
\caption{Comparison of the frational EW variations of \ion{C}{iv} BALs and those of \ion{Si}{iv} BALs. The black dashed line indicates equal strengths of the frational EW variations for the two ions. The red solid line is the Bayesian linear regression fit} for all data points. The small frame in the lower right shows a clear view without err bars of the same plots.
    \label{fig.2}
\end{figure}
In this Letter, we attempt to test whether the variation of BALs and that of the continuum are correlated, based on a sample of multi-observed SDSS quasars, which possesses variable \ion{C}{iv} and \ion{Si}{iv} BALs. This Letter is organized as follows. We describe data selection and preparation in Section 2, and present data analysis in  Section 3. In Section 4 we make a discussion about our results. A summary is given in Section 5.

\begin{figure*}
    \includegraphics[width=\columnwidth]{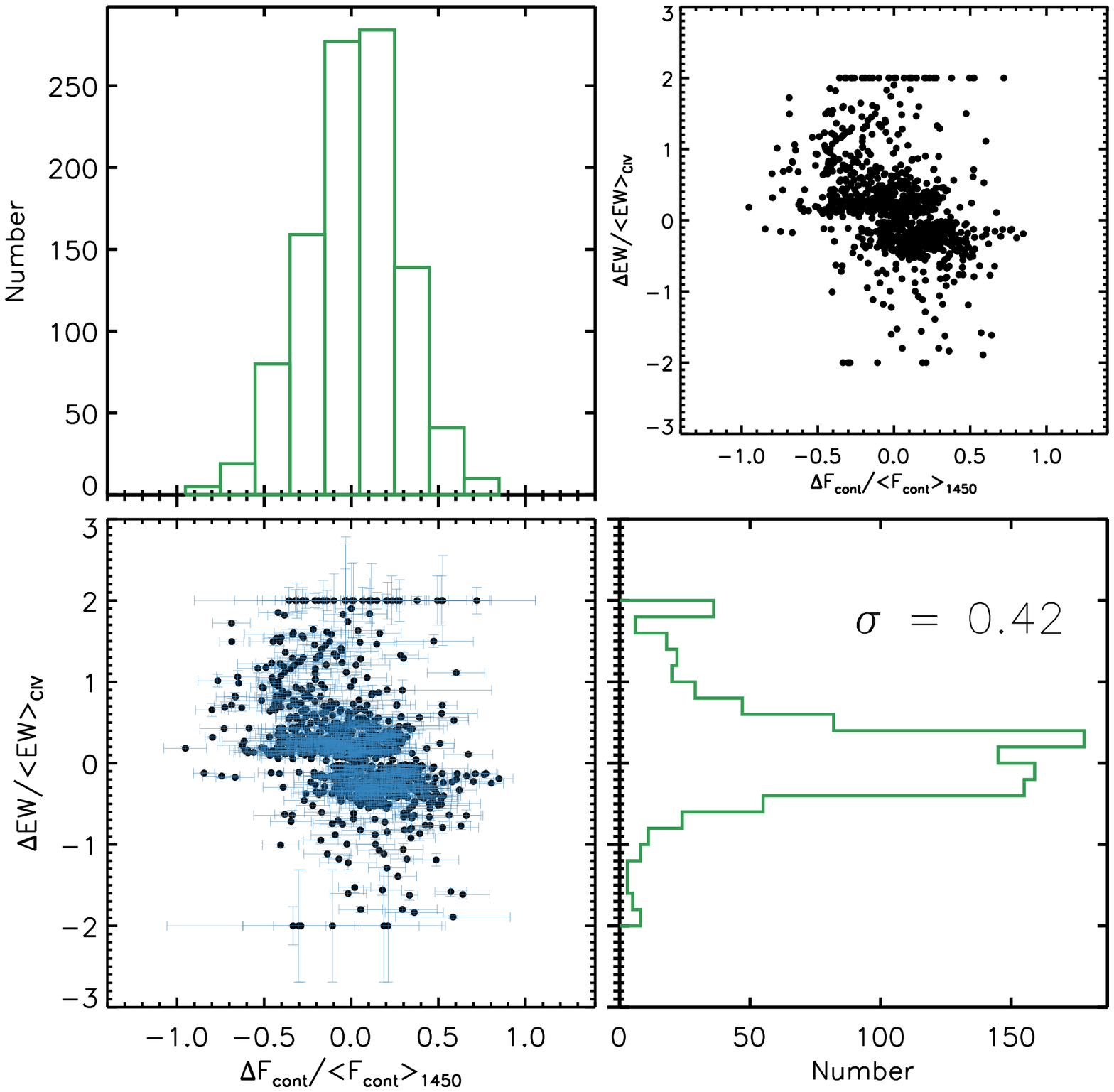}
    \includegraphics[width=\columnwidth]{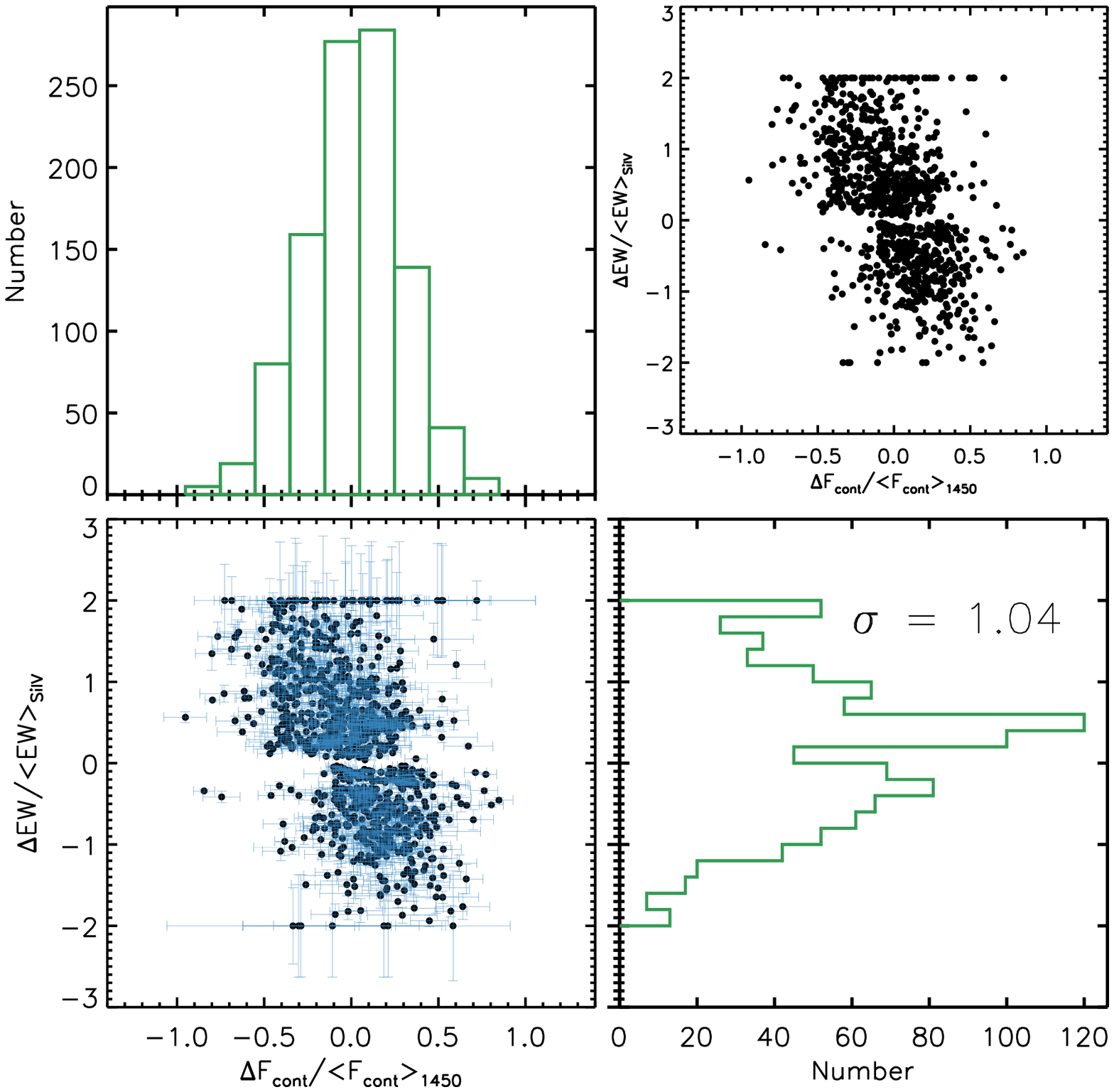}
 \caption{Plots of fractional EW variations of BALs and fractional flux variations of the continuum (four left-hand panels: $\rm \Delta EW/\langle EW \rangle_{\ion{C}{iv}}$ versus $\Delta \rm F_{\rm cont}/\langle \rm F_{cont} \rangle_{1450}$; four right-hand panels: $\rm \Delta EW/\langle EW \rangle_{\ion{Si}{iv}}$ versus $\Delta \rm F_{\rm cont}/\langle \rm F_{cont} \rangle_{1450}$). Histograms show number distributions of $\Delta \rm F_{\rm cont}/\langle \rm F_{cont} \rangle_{1450}$ (top axes) and those of $\Delta \rm EW/\langle EW \rangle$ (right-hand axes). The values of $\sigma$ of the Gaussian components that fit the number distributions of $\rm \Delta EW/\langle EW \rangle_{\ion{C}{iv}}$ and $\rm \Delta EW/\langle EW \rangle_{\ion{Si}{iv}}$ are labelled. The median error values for $\rm \Delta EW/\langle EW \rangle_{\ion{C}{iv}}$, $\rm \Delta EW/\langle EW \rangle_{\ion{Si}{iv}}$, and $\Delta \rm F_{\rm cont}/\langle \rm F_{cont} \rangle_{1450}$ are 0.026, 0.067 and 0.086, respectively. The frame in the top right-hand side of each panel shows a clear view without error bars.}
    \label{fig.3}
\end{figure*}

\begin{figure*}
\includegraphics[width=2\columnwidth]{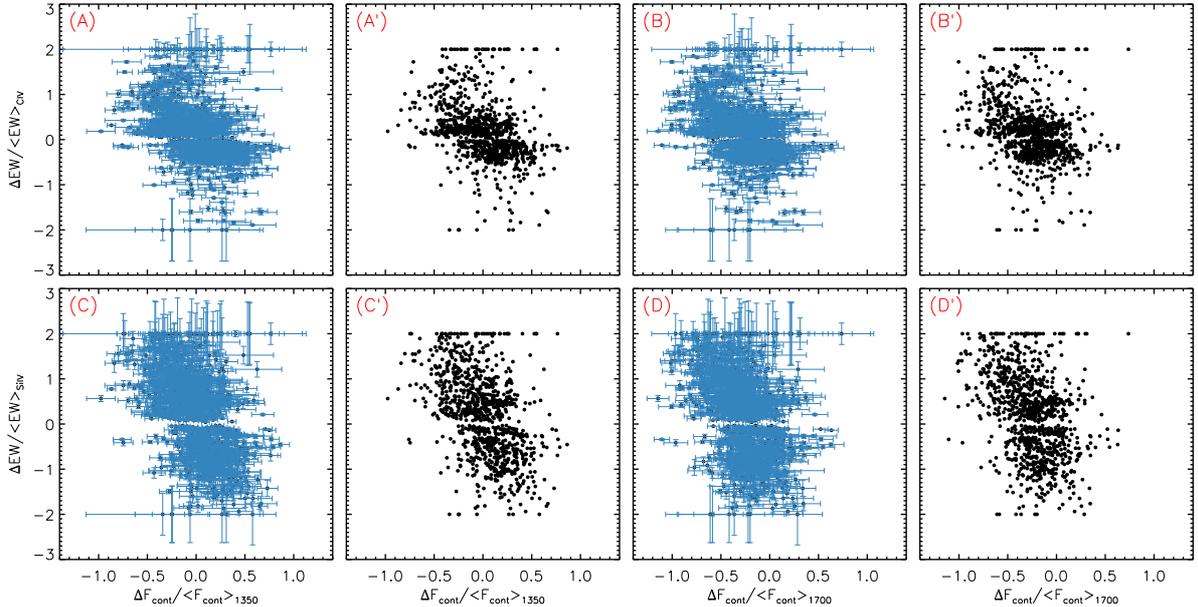}
\caption{Fractional EW variations of \ion{C}{iv} (upper panels) and \ion{Si}{iv} (lower panels) BAL troughs as a function of fractional flux variations of the continuum (four left-hand panels are for $\Delta \rm F_{\rm cont}/\langle \rm F_{cont} \rangle$ measured at 1350 \AA; four right-hand panels are for $\Delta \rm F_{\rm cont}/\langle \rm F_{cont} \rangle$ measured at 1700 \AA). Panels (A\arcmin), (B\arcmin), (C\arcmin) and (D\arcmin) are the clear versions without error bars for Panels (A), (B), (C) and (D). The median error values for $\Delta \rm F_{\rm cont}/\langle \rm F_{cont} \rangle_{1350}$ and $\Delta \rm F_{\rm cont}/\langle \rm F_{cont} \rangle_{1700}$ are 0.096 and 0.098, respectively.}
    \label{fig.4}
\end{figure*}

\section{Sample selection and preparation}
He et al. (2017) presented a BAL quasar catalog\footnote{\url{http://home.ustc.edu.cn/~zcho/SDSS\_DR12\_BAL/}} that contains 2005 BAL quasars, which were observed twice or more times by SDSS-I/II/III. He et al. (2017) selected these quasars according to two criteria: (1) a redshift range of $1.9 \textless z \textless 4.7$; (2) a signal-to-noise ratio (S/N) level of S/N $\textgreater$10 in at least one spectrum. First, we did a selection with these spectra according to the differences in EW of \ion{C}{iv} or \ion{Si}{iv} BALs between the two SDSS observations, $\rm \Delta W$, which is defined as follows:

\begin{equation}
	    \frac{\rm \Delta W}{\sigma'_{\rm w}}= \frac{\rm W_{2}-W_{1}}{\sqrt{\sigma_{\rm w1}^2+\sigma_{\rm w2}^2}},
    \label{eq.1}
\end{equation}
where $\rm W_{1}$ and $\rm W_{2}$ represent the EWs of \ion{C}{iv} or \ion{Si}{iv} BALs measured from two-epoch spectra of a quasar, $\rm \sigma_{w1}$ and $\sigma_{w2}$ represent the errors on $\rm W_{1}$ and $\rm W_{2}$, respectively. We only keep spectrum pairs that with a confident level of $\rm \Delta W\textgreater 5\sigma'_{w}$ for both \ion{C}{iv} and \ion{Si}{iv} lines, thus reduced the sample to 1014 spectrum pairs in 483 quasars. Then, we downloaded these spectra from SDSS data release 14 (DR14; Abolfathi et al. 2017). Since the flux calibration of SDSS DR14 has been improved (Abolfathi et al. 2017), we can get more accurate values of the flux density.

In order to evaluate variation of the continuum, we fitted each spectrum using a power-law function. The power-law continuum was fitted iteratively in several relatively line-free (RLF) wavelength regions (1250--1350, 1700--1800, 1950--2200, 2650--2710 \AA~in the rest-frame), which were defined by Gibson et al. (2009). During the fitting, we masked out the pixels that are beyond 3$\sigma$ significance to reduce the influences of absorption/emission lines and remaining sky pixels. An example of the power-law continuum fit is shown in Fig. \ref{fig.1}.
\begin{table*}
    \centering
    \caption{Summary of the correlation coefficients.}
    \label{tab.1}
    \begin{tabular}{lrrrrr} 
        \hline\noalign{\smallskip}
	& \makecell[c]{$r(p)^{\rm a}$} &  $\makecell[c]{r_{\rm Bayes}^{\rm b}}$  & \makecell[c]{$k_{\rm Bayes}^{\rm c}$}  & \makecell[c]{$b_{\rm Bayes}^{\rm d}$} & \makecell[c]{$\sigma_{\rm int}^{\rm e}$}\\
        \hline\noalign{\smallskip}
$\rm \Delta F_{cont}/\langle F_{cont} \rangle _{1450}-\Delta EW/\langle EW \rangle_{\ion{C}{iv}}$&	--0.43 (\textless1E--44)	&	--0.47 	$\pm$	0.03 	&	--1.14 	$\pm$	0.08 	&	0.17 	$\pm$	0.02 	&	0.29 	\\
$\rm \Delta F_{cont}/\langle F_{cont} \rangle _{1450}-\Delta EW/\langle EW \rangle_{\ion{Si}{iv}}$	&	--0.48 (\textless1E--44)	&	--0.57 	$\pm$	0.02 	&	--1.93 	$\pm$	0.10 	&	0.20 	$\pm$	0.02 	&	0.51 	\\
$\rm \Delta F_{cont}/\langle F_{cont} \rangle _{1350}-\Delta EW/\langle EW \rangle_{\ion{C}{iv}}$	&	--0.45 (\textless1E--44)	&	--0.50 	$\pm$	0.03 	&	--1.16 	$\pm$	0.08 	&	0.16 	$\pm$	0.02 	&	0.28 	\\
$\rm \Delta F_{cont}/\langle F_{cont} \rangle _{1350}-\Delta EW/\langle EW \rangle_{\ion{Si}{iv}}$	&	--0.49 (\textless1E--44)	&	--0.59 	$\pm$	0.02 	&	--1.94 	$\pm$	0.10 	&	0.18 	$\pm$	0.02 	&	0.49 	\\
$\rm \Delta F_{cont}/\langle F_{cont} \rangle _{1700}-\Delta EW/\langle EW \rangle_{\ion{C}{iv}}$	&	--0.34 (1.33E--28)	&	--0.43 	$\pm$	0.03 	&	--1.06 	$\pm$	0.09 	&	--0.12 	$\pm$	0.03 	&	0.31 	\\
$\rm \Delta F_{cont}/\langle F_{cont} \rangle _{1700}-\Delta EW/\langle EW \rangle_{\ion{Si}{iv}}$&	--0.41 (5.59E--42)	&	--0.51 	$\pm$	0.03 	&	--1.81 	$\pm$	0.11 	&	--0.30 	$\pm$	0.04 	&	0.55 	\\
$\rm \Delta EW/\langle EW \rangle_{\ion{C}{iv}}-\Delta EW/\langle EW \rangle_{\ion{Si}{iv}}$&	 0.95 (\textless1E--44)	&	0.93 	$\pm$	0.00 	&	1.39 	$\pm$	0.02 	&	--0.02 	$\pm$	0.01 	&	0.11 	\\
        \hline
    \end{tabular}
 \begin{tablenotes}
  \footnotesize
  \item \emph{Notes.}$^{\rm a}$The Spearman rank correlation coefficient, the values in brackets are the two-sided significance of the deviation from zero.
  \item $^{\rm b}$The correlation coefficient between the dependent and independent variables from Bayesian linear regression fit . 
  \item $^{\rm c}$The slope of the linear fit.
  \item $^{\rm d}$The constant in the regression.
  \item $^{\rm e}$The variance of the intrinsic scatter. 
  \end{tablenotes}
\end{table*}
\section{data analysis}
To calculate fractional EW variations for \ion{C}{iv} BALs and \ion{Si}{iv} BALs and corresponding uncertainties, we use the following equations:  
\begin{equation}
\frac{\Delta \rm EW}{\langle \rm EW \rangle}=\frac{\rm EW_2-\rm EW_1}{(\rm EW_2 + \rm EW_1)\times 0.5},
    \label{eq.EW}
\end{equation}

\begin{equation}
\sigma_{\frac{\Delta \rm EW}{\langle \rm EW \rangle}}=\frac{4\sqrt{\rm EW_{\rm 2}^2\rm EW_{\rm noise1}^2+\rm EW_{\rm 1}^2\rm EW_{\rm noise2}^2}}{(\rm EW_{\rm 2} + \rm EW_{\rm 1})^2},
    \label{eq.eEW}
\end{equation}
where the subscripts `1' and `2' are the former and latter time epoch measurements, respectively; $\langle \rm EW \rangle$ is the average EW value for the two epochs; $\rm EW_{\rm noise}$ is the EW uncertainty. The EW measurements of \ion{C}{iv} and \ion{Si}{iv} BALs were derived from He et al. (2017). The $\Delta \rm EW/\langle EW \rangle_\ion{C}{iv}$ versus $\Delta \rm EW/\langle EW \rangle_\ion{Si}{iv}$ correlation were plotted in Fig. \ref{fig.2}.

Similarly, we evaluated fractional flux variation of the continuum and  corresponding uncertainty using the following equations:
\begin{equation}
\frac{\Delta \rm F_{\rm cont}}{\langle \rm F_{cont} \rangle}=\frac{\rm F_{\rm cont2}-\rm F_{\rm cont1}}{(\rm F_{\rm cont2} + \rm F_{\rm cont1})\times 0.5},
    \label{eq.F}
\end{equation}
\begin{equation}
\sigma_{\frac{\Delta \rm F_{\rm cont}}{\langle \rm F_{cont} \rangle}}=\frac{4\sqrt{\rm F_{\rm cont2}^2\rm F_{\rm noise1}^2+\rm F_{\rm cont1}^2\rm F_{\rm noise2}^2}}{(\rm F_{\rm cont2} + \rm F_{\rm cont1})^2}
    \label{eq.eF},
\end{equation}
where $\rm F_{\rm cont1}$ and $\rm F_{\rm cont2}$ represent the power-law continuum flux for the two epochs, respectively; $\rm F_{\rm noise}$ is the flux uncertainty.
The wavelength coverage of SDSS-I/II spectra is 3800--9200 \AA~(observed frame), and the emission redshift coverage of our spectral sample is $1.9 \textless z \textless 4.0$. Thus, the common wavelength coverage of our spectral sample is about 1305--1840 \AA~in rest frame. To avoid accidental conclusions, we measured $\Delta \rm F_{\rm cont}/\langle \rm F_{cont} \rangle$ at 1350 ($\Delta \rm F_{\rm cont}/\langle \rm F_{cont} \rangle_{1350}$), 1450 ($\Delta \rm F_{\rm cont}/\langle \rm F_{cont} \rangle_{1450}$) and 1700 \AA~($\Delta \rm F_{\rm cont}/\langle \rm F_{cont} \rangle_{1700}$), respectively. 

To quantitatively assess whether there is a correlation between the variation of BALs and that of the continuum, we performed a Spearman correlation and a Bayesian linear regression techniques (Kelly 2007). The test results were listed in Table \ref{tab.1}. 
  
\section{results and Discussion}
\subsection{Correlation between fractional variation of BALs and that of the continuum}
Both the Spearman and Bayesian tests show moderate anticorrelations with high significance level between fractional variations of BALs and that of the continuum, for both \ion{C}{iv} and \ion{Si}{iv} BALs, and for $\rm \Delta F_{\rm cont}/\langle \rm F_{cont} \rangle_{1350}$, $\rm \Delta F_{\rm cont}/\langle \rm F_{cont} \rangle_{1450}$ and $\rm \Delta F_{\rm cont}/\langle \rm F_{cont} \rangle_{1700}$ (Table \ref{tab.1}, Fig. \ref{fig.3} and Fig. \ref{fig.4}). To make the statement clear, we select the $\rm \Delta F_{\rm cont}/\langle \rm F_{cont} \rangle_{1450}$ to present the fractional variation of the continuum in the following. The Spearman coefficient is $r=-0.43$ (with a significance level of $p \textless1\rm E-44$) for the $\Delta \rm EW/\langle EW \rangle_\ion{C}{iv}$ versus $\rm \Delta F_{\rm cont}/\langle \rm F_{cont} \rangle_{1450}$ correlation and $r=-0.48~(p \textless1\rm E-44$) for the $\rm \Delta F_{\rm cont}/\langle \rm F_{cont} \rangle_{1450}$ versus $\Delta \rm EW/\langle EW \rangle_\ion{Si}{iv}$ correlation. The correlation also be confirmed by the Bayesian linear regression fit, the method that have taken into account the intrinsic scatter in the data. These results imply that the variations of these BAL variabilities are primarily driven by the variation of an ionizing continuum.

Although correlations with a high significance significance level do exist, both the two relations ($\Delta \rm EW/\langle EW \rangle_\ion{C}{iv}$ versus $\rm \Delta F_{\rm cont}/\langle \rm F_{cont} \rangle_{1450}$ and $\Delta \rm EW/\langle EW \rangle_\ion{Si}{iv}$ versus $\rm \Delta F_{\rm cont}/\langle \rm F_{cont} \rangle_{1450}$) show large dispersions (Fig. \ref{fig.3}). We suppose that such dispersions may be caused by several reasons. (1)The continuum may be effected by variable shielding gas, which prevents us from measuring accurately the $\rm \Delta F_{\rm cont}/\langle \rm F_{cont} \rangle_{1450}$ (Kaastra et al. 2014; Arav et al. 2015). (2) BALs may suffer from saturation, which prevents us from measuring accurately the $\Delta \rm EW/\langle EW \rangle_\ion{C}{iv}$ and $\Delta \rm EW/\langle EW \rangle_\ion{Si}{iv}$. (3)The existence of other mechanisms leading to vairations of the BALs, for instance, gas moving in or out of the line of sight may at work for some of the BALs. (4)The variaion of a BAL trough may respond to a variation of continuum positively or negatively, depending on the ionization states of absorbers (Wang et al. 2015; He et al. 2017).

\subsection{Comparison of fractional EW variations of C IV and Si IV BALs}
As shown in Fig. \ref{fig.2} and Table \ref{tab.1}, correlation between fractional EW variation of \ion{C}{iv}($\Delta \rm EW/\langle EW \rangle_\ion{C}{iv}$) and those of \ion{Si}{iv} ($\Delta \rm EW/\langle EW \rangle_\ion{Si}{iv}$) are detected ($r=0.95, p\textless\rm 1E-44$). This result is in agreement with previous studies (Filiz Ak et al. 2013; Wildy et al. 2014; He et al. 2017). However, the slope value of $\Delta \rm EW/\langle EW \rangle_\ion{C}{iv}$ versus $\Delta \rm EW/\langle EW \rangle_\ion{Si}{iv}$ relation is greater than 1 ($k_{\rm Bayes}=1.39$), which means that \ion{Si}{iv} BALs show larger fractional EW variations than \ion{C}{iv} BALs. This difference between these two ions is also shown in Fig. \ref{fig.3}, where $\Delta \rm EW_\ion{Si}{iv}$ (with $\sigma = 1.04$ for the best-fitting Gaussian component) tends to be greater than $\Delta \rm EW_\ion{C}{iv}$ (with $\sigma = 0.42$ for the best-fitting Gaussian component). We propose that this difference between the two ions is caused primarily by the following reasons. First, the red and blue lines of $\ion{Si}{iv}\lambda\lambda1393,1402$ doublets are separated wider than those of $\ion{C}{iv}\lambda\lambda1548,1551$ doublets in the rest frame, due to differences in their fine structures. Second, as shown in both photoionzation simulation  (e.g., He et al. 2017) and observation studies (e.g., Filiz Ak et al. 2013; He et al. 2017), \ion{C}{iv} BAL troughs tend to be stronger than \ion{Si}{iv} BAL troughs. These two factors cause the \ion{C}{iv} BALs suffer more saturations than \ion{Si}{iv} BALs, so that $\Delta \rm EW/\langle EW \rangle_\ion{C}{iv}$ tends to be weaker than $\Delta \rm EW/\langle EW \rangle_\ion{Si}{iv}$.

\subsection{Other properties of the absorbers}
Our research can roughly reveal the ionization states of the absorbers. Photoionization simulations have showed that with increasing ionization parameter (U), the EWs of $\ion{C}{IV}$ and $\ion{Si}{IV}$ rise first, then reach a peak, and decrease at last (e.g., He et al. 2017). In this Letter, fractional EW variations of both \ion{C}{iv} and \ion{Si}{iv} show anticorrelations with fractional flux variations of the continuum. These results mean that most of the absorbers in our sample are at the relatively high ionization state.   

Recently, Lu et al. (2017) have found a strong anticorrelation between the variations of NALs and that of the continuum, using a sample of two-epoch optical spectra of 40 quasars with 52 variable \ion{C}{iv} $\lambda \lambda$1548,1551 doublets (the sample is from Chen et al. 2015). Now in this Letter, we find out the anticorrelation between the variations of BALs and the continuum too. The variations of both BALs and NALs mainly driven by photoionization, indicating that there may be some physical relationship between the variable BALs and NALs. Specifically, they may origin from the same part of the outflow having similar locations (e.g., close to the central engine) and physical conditions (e.g., high ionization state).

\section{SUMMARY}
Utilizing a sample of 1014 spectrum pairs in 483 quasars, we have calculated fractional variations of the continuum, $\rm \Delta F_{\rm cont}/\langle \rm F_{cont} \rangle_{1450}$, based on the power-law continuum fitted by ourselves; and fractional variations for \ion{C}{iv} and \ion{Si}{iv} BALs, $\Delta \rm EW/\langle EW \rangle_\ion{C}{iv}$ and $\Delta \rm EW/\langle EW \rangle_\ion{Si}{iv}$, based on the EW measurements from He et al. (2017). And we have probed the correlations  between the three. The main conclusions can be summarized as follows.

\begin{enumerate}
 \item Moderate anticorrelations with high significance level of $\rm \Delta F_{\rm cont}/\langle \rm F_{cont} \rangle_{1450}$ versus $\Delta \rm EW/\langle EW \rangle_\ion{C}{iv}$ and $\rm \Delta F_{\rm cont}/\langle \rm F_{cont} \rangle_{1450}$ versus $\Delta \rm EW/\langle EW \rangle_\ion{Si}{iv}$ are detected. These findings provide evidence for photoionization-driven BAL variations. However, the correlations show large dispersions, which can be explained by the effects from variable shielding gas, BAL saturation, ionization state of the outflow or other BAL variation machines.
 \item  Significant positive correlation between fractional EW variation of \ion{C}{iv} ($\Delta \rm EW/\langle EW \rangle_\ion{C}{iv}$) and those of \ion{Si}{iv} ($\Delta \rm EW/\langle EW \rangle_\ion{Si}{iv}$) is detected (Spearman coefficient $r=0.95$ with $p \textless \rm 1E-44$). However, the variations of \ion{Si}{iv} BALs tend to be greater than those of \ion{C}{iv} BALs, which can be explained by the difference in saturation extent between \ion{C}{iv} and \ion{Si}{iv} BALs.

\end{enumerate}

\section*{Acknowledgements}
The authors acknowledge the anonymous referee for helpful comments that
contributed to the betterment of this Letter. In addition, we thank He et al. for making the multi-epoch BAL catalog publicly available. 

Funding for SDSS-III was provided by the Alfred P. Sloan Foundation, the
Participating Institutions, the National Science Foundation and the US
Department of Energy Office of Science. The SDSS-III web site is
\url{http://www.sdss3.org/.}

SDSS-III is managed by the Astrophysical Research Consortium for the
Participating Institutions of the SDSS-III Collaboration, including the
University of Arizona, the Brazilian Participation Group, Brookhaven National Laboratory, Carnegie Mellon University, University of Florida, the French Participation Group, the German Participation Group, Harvard University, the Instituto de Astrofisica de Canarias, the Michigan State/Notre Dame/JINA Participation Group, Johns Hopkins University, Lawrence Berkeley National Laboratory, Max Planck Institute for Astrophysics, Max Planck Institute for Extraterrestrial Physics, New Mexico State University, New York University, Ohio State University, Pennsylvania State University, University of Portsmouth, Princeton University, the Spanish Participation Group, University of Tokyo, University of Utah, Vanderbilt University, University of Virginia, University of
Washington, and Yale University.





\begin{thebibliography}{99}
\bibitem[\protect\citeauthoryear{}{2015}]{20}
Abolfathi B. et al., 2017, preprint (arXiv:1707.09322)
\bibitem[\protect\citeauthoryear{}{2015}]{16}
Arav N. et al., 2015, A\&A, 577, 37
\bibitem[\protect\citeauthoryear{}{}]{3}
Barlow T. A., Junkkarinen V. T., Burbidge E. M., 1989, ApJ, 347, 674
\bibitem[\protect\citeauthoryear{}{}]{6}
Chen Z., Gu Q., Chen Y., Cao Y., 2015, MNRAS, 450, 3904
\bibitem[\protect\citeauthoryear{}{}]{6}
Crenshaw D. M., Kraemer S. B., George I. M., 2003, ARA\&A, 41, 117
\bibitem[\protect\citeauthoryear{}{2013}]{10}
Filiz Ak N. et al., 2013, ApJ, 777, 168
\bibitem[\protect\citeauthoryear{}{}]{1}
Foltz C. B., Weymann R. J., Morris S. L., Turnshek D. A., 1987, ApJ, 317, 450 
\bibitem[\protect\citeauthoryear{}{2008}]{7}
Gibson R. R., Brandt W. N., Schneider D. P., Gallagher S. C., 2008, ApJ, 675, 985
\bibitem[\protect\citeauthoryear{}{2008}]{21}
Gibson R. R. et al., 2009, ApJ, 692, 758
\bibitem[\protect\citeauthoryear{}{2008}]{25}
Hamann F., Kaplan K. F., Rodr\'iguez Hidalgo P., Prochaska J. X., Herbert-Fort S., 2008, MNRAS, 391, L39
\bibitem[\protect\citeauthoryear{}{2017}]{12}
He Z., Wang T., Zhou H., Bian W., Liu G., Yang C., Dou L., Sun L., 2017, APJS, 229, 22
\bibitem[\protect\citeauthoryear{}{}]{15}
Kaastra J. S. et al., 2014, Science, 345, 64
\bibitem[\protect\citeauthoryear{}{}]{22}
Kelly B. C., 2007, ApJ, 665, 1489
\bibitem[\protect\citeauthoryear{}{Paper I}]{14}
Lu W. et al., 2017, MNRAS, 468, L6
\bibitem[\protect\citeauthoryear{}{}]{13}
P\^aris I. et al., 2017, A\&A, 597, A79
\bibitem[\protect\citeauthoryear{}{2016}]{5}
Shen Y. et al., 2011, ApJS, 194, 45
\bibitem[\protect\citeauthoryear{}{2016}]{5}
Shi X. et al., 2016, ApJ, 819, 99
\bibitem[\protect\citeauthoryear{}{}]{2}
Smith L. J., Penston M. V., 1988, MNRAS, 235, 551
\bibitem[\protect\citeauthoryear{}{2014}]{9}
Vivek M., Srianand R., Petitjean P., Mohan V., Mahabal A., Samui S., 2014, MNRAS, 440, 799
\bibitem[\protect\citeauthoryear{}{1991}]{36}
Weymann R. J., Morris S. L., Foltz C. B., Hewett P. C., 1991, ApJ, 373, 23
\bibitem[\protect\citeauthoryear{}{2014}]{8}
Wildy C., Goad M. R., Allen J. T., 2014, MNRAS, 437, 1976
\bibitem[\protect\citeauthoryear{}{2015}]{11}
Wang T., Yang C., Wang H., Ferland G., 2015, ApJ, 814, 150
\end{thebibliography}






\bsp    
\label{lastpage}
\end{document}